\documentclass[twoside]{article}

\usepackage[accepted]{aistats2020_arxiv}
\usepackage[round]{natbib}

\usepackage{booktabs}

\usepackage{color, xcolor}
\usepackage{latexsym}              
\usepackage{amsmath}               
\usepackage{amssymb}               
\usepackage{amsfonts}              
\usepackage{amsthm}                
\usepackage{multirow}
\usepackage{tikz}                  
\usetikzlibrary{arrows,positioning,shapes}
\usepackage{tcolorbox}
\usepackage{IEEEtrantools}
\usepackage{appendix}
\usepackage{algorithm}
\usepackage[noend]{algpseudocode}

\usepackage{multicol, blindtext}

\newcommand{\figref}[1]{Figure~\ref{#1}}
\newcommand{\secref}[1]{Section~\ref{#1}}

\usepackage{xspace}
\usepackage[nolist,smaller]{acronym}
\newcommand{\acro}[1]{\textsc{#1}\xspace}
\newcommand{\TVD}{\acro{\smaller TVD}}
\newcommand{\KLD}{\acro{\smaller KLD}}
\newcommand{\NPL}{\acro{\smaller NPL}}
\newcommand{\DP}{\acro{\smaller DP}}
\newcommand{\argmin}{\operatorname{argmin}}
 
\newtheorem{proposition}{Proposition} 
\newtheorem{corollary}{Corollary}
\newtheorem{assumption}{Assumption}

\definecolor{TVDColor}{HTML}{009E73}
\definecolor{KLDColor}{HTML}{0072B2}

\newenvironment{proofsketch}{%
\proof}{\endproof}

\usepackage{xspace}
\newcommand{\phat}[1][]{\ensuremath{\widehat{p}^{\, #1}}\xspace}

\begin{document}

%

%

\twocolumn[

\aistatstitle{Robust Bayesian Inference for Discrete Outcomes with the Total Variation Distance}

\aistatsauthor{ Jeremias Knoblauch$^\ast$ \And Lara Vomfell$^\ast$ }

\aistatsaddress{ University of Warwick \& The Alan Turing Institute \And  University of Warwick } ]

\begin{abstract}
Models of discrete-valued outcomes are easily misspecified if the data exhibit zero-inflation, overdispersion or contamination. Without additional knowledge about the existence and nature of this misspecification, model inference and prediction are adversely affected. Here, we introduce a robust discrepancy-based Bayesian approach using the Total Variation Distance (\TVD).
  In the process, we address and resolve two challenges: 
  First, we study convergence and robustness properties of a computationally efficient estimator for the \TVD between a parametric model and the data-generating mechanism. 
Secondly, we provide an efficient inference method adapted from \citet{LyddonBayesBootstrap} which corresponds to formulating an uninformative nonparametric prior directly over the data-generating mechanism.
  Lastly, we empirically demonstrate that our approach is robust and significantly improves predictive performance on a range of simulated and real world data. 
\end{abstract}

\begin{figure}[b!]
    \centering
    \includegraphics[width=\columnwidth, trim= {0.0cm 0.4cm 0.cm 0.3cm}, clip]{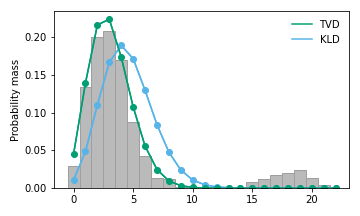}
    \caption{Discrete count data with some outliers (grey histogram) are modelled with a Poisson distribution. 
    Inferring the model with a standard Bayesian approach amounts to minimizing the \textcolor{KLDColor}{\textbf{Kullback-Leibler Divergence (\KLD)}} between model and data. 
    For this case, the outliers have a disproportionate impact.
    Minimizing the \textcolor{TVDColor}{\textbf{Total Variation Distance (\TVD)}} instead is robust to such contamination.
    }
    \label{fig:toy}
\end{figure}

\section{INTRODUCTION}
Discrete outcomes such as counts or classification labels pose significant modelling challenges because standard inference is vulnerable to over-weighting subtle data features such as boundary or censoring effects,  zero-inflation, as well as issues such as outliers, inliers and corrupted data.

Building on a growing literature on robustness in Bayesian models, the current paper provides a generic strategy for robustness in this setting: 
In the absence of more detailed knowledge on the nature of misspecification which could then be explicitly modelled, we ensure robustness implicitly via the learning criterion.
As outlined in \cite{Jewson}, a particularly appealing way of doing so is by way of a generalized Bayesian approach \citep{Bissiri} based on robust discrepancies.
In the context of continuous data, this idea was first pioneered using $\alpha$-divergences \citep{MinDisparities} and has since been extended to $\beta$- and $\gamma$-divergences
\citep{GoshBasuPseudoPosterior, AISTATSBetaDiv, RBOCPD, GVI,  betaDivCores}
as well the Maximum Mean Discrepancy \citep{BadrMMD}. 
For various reasons, all these approaches are somewhat unattractive for statistical machine learning problems with discrete-valued data: The approach of \citet{MinDisparities} relies on a computationally inefficient kernel density estimate, $\beta$- and $\gamma$-divergences have no easily computable form outside the exponential family, and the Maximum Mean Discrepancy relies on kernels in a way that makes it less obvious how to work with it in discrete settings.

In light of these limitations and as illustrated in Figure \ref{fig:toy}, we propose a generalized Bayesian inference method for discrete-valued outcomes based on the Total Variation Distance (\TVD).
We make three contributions:
\begin{enumerate}
    \item We explore the theoretical properties of our estimator for the \TVD and find that it  satisfies exponential concentration inequalities (Propositions \ref{proposition:concentration_discrete} and \ref{proposition:concentration_continuous}), converges almost surely (Corollary \ref{corollary:as_convergence}) and retains the robustness properties of the true \TVD (Corollary \ref{corollary:robustness_estimator}). 
    Further, the estimator's minimizer is  strongly consistent (Proposition \ref{proposition:consistency}).
    %
    %
    \item We adapt Nonparametric Bayesian Learning (\NPL) as popularized by \cite{SimonNPL} and \cite{EdwinNPL} to our setting. As the resulting algorithm is computationally equivalent to the Bayesian Bootstrap, inference has low computational complexity and is embarassingly parallel.
    \item We apply the new inference scheme to a range of simulated and real world datasets.  As expected, the \TVD yields superior performance under misspecification. Even in the absence of misspecification, we match performance of the \KLD. 
\end{enumerate}

In Section \ref{sec:divergences_and_inference}, we briefly recap  divergence-based generalizations of Bayesian inference. 
Next, Section \ref{sec:motivation} motivates the use of the \TVD within this framework. 
As the \TVD needs to be estimated, we prove a number of properties exhibited by our estimator in Section \ref{sec:estimator_properties}.
Section \ref{sec:nonparametric_learning} explains how to embed out estimator into Nonparametric Bayesian Learning (\NPL). 
We then apply the resulting algorithm to a range of datasets and discuss the results in Section \ref{sec:experiments}.
We conclude that the method constitutes a reliable and generic robustness strategy for inference in discrete outcome models.

\section{DIVERGENCES \& INFERENCE}
\label{sec:divergences_and_inference}

It is well-known that both Maximum Likelihood estimation and conventional Bayesian Inference minimize the Kullback-Leibler divergence (\KLD) between the empirical density $\phat_n$ and a model family $\{f_{\theta}: \theta \in \Theta\}$.
The following paragraphs introduces these ideas and explains the role of the \TVD in the current paper.

\subsection{Disparity-minimizing Estimators}
The Maximum Likelihood objective is given by
%
\begin{IEEEeqnarray}{rCl}
     L_n(\theta) & = & \frac{1}{n}\sum_{i=1}^n \log f_{\theta}(y_i|x_i) 
     \nonumber \\
     & = & 
     \mathbb{E}_{{\widehat{p}}_n^x}\Bigg[ 
     \underbrace{
     \mathbb{E}_{\phat[y|x]_n}\!\!\left[ 
         \log\! \left(
            \frac{ f_{\theta}(y|x)}{\phat_n(y|x)} 
        \right) 
     \right]
     }_{ = - \KLD(\phat[y|x]_n\|f_{\theta}) }
     \Bigg]
     -
     H(\widehat{p}_n)
     ,
     \label{eq:ML=KLD}
\end{IEEEeqnarray}

where $H$ denotes the Shannon-entropy, and $\phat_n(y,x) = n^{-1}\sum_{i=1}^n\delta_{(x_i, y_i)}(x,y)$, $\phat[x]_n(x) = n^{-1}\sum_{i=1}^n\delta_{x_i}(x)$ and $\phat[y|x]_n(y|x) = \phat_n(y,x)/\phat[x]_n(x)$ define the joint, marginal and conditional empirical distributions of a sample $\{y_i, x_i\}_{i=1}^n$.
As $H(\phat_n)$ does not depend on $\theta$, \eqref{eq:ML=KLD} shows that maximizing $L_n$ over $\theta$ amounts to minimizing $\mathbb{E}_{\phat[x]_n}[\KLD(\phat[y|x]_n\|f_{\theta})]$ over $\theta$.
Because the \KLD is not robust to outliers and misspecification, this observation has inspired numerous alternative disparity-based techniques. While these methods were initially focused on statistical testing procedures \citep{MinimumDistanceMethod}, they were quickly extended in order to derive robust estimators \citep[e.g.][]{ MinimumHellingerEstimator, TVDestimate1985, MinimumHellingerEstimatorCountData, BasuDPD, scoreMatching, FXMMD, MinimumSteinEstimator, MMDFiniteSample}.

\subsection{Bayesian inference}

Defining $\pi(\theta)$ as the prior distribution, one can rewrite the Bayesian posterior given by $q_n(\theta) \propto \prod_{i=1}^n f_\theta(y_i|x_i)\pi(\theta)$ in similar fashion.
Specifically, $q_n$ solves a well-known variational problem: For $\mathcal{P}(\Theta)$ denoting the set of all probability measures on $\Theta$, 
\begin{IEEEeqnarray}{rCl}
    q_n(\theta) = \underset{q \in \mathcal{P}(\Theta)}{\argmin}\left\{
        \mathbb{E}_{q}\left[
          \KLD(\widehat{p}_n\|f_{\theta})
        \right]
        +
        \frac{1}{n}\KLD(q\|\pi)
    \right\}.
    \nonumber
\end{IEEEeqnarray}
Notice that this Bayesian objective is equivalent to a prior-regularized version of its Frequentist counterpart.
This perspective on Bayesian inference is particularly popular within the PAC-Bayesian and Variational Inference communities \citep[see e.g.][]{PACmeetsBayesianInference,GVI,PACPrimer}.
In fact, the similarities with the frequentist version of the objective ensures that their solutions coincide as $n\to\infty$ \citep[see e.g.][]{BayesConsistencyReview}, even if the model is misspecified \citep[e.g.][]{BayesianConsistencyNonIID}, even if the discrepancy term assessing the model fit is no longer the \KLD \citep[e.g.][]{GoshBasuPseudoPosterior, BayesianConsistencyMiller}, and even if the prior regularization term is no longer the \KLD \citep{GVIConsistency}.

It is important to note this asymptotic equivalence: It implies that model misspecification issues plaguing Frequentist estimators for $\theta$ will carry over into Bayesian inference on $\theta$ if $n$ is large enough.

\section{MOTIVATION}
\label{sec:motivation}

To address such robustness concerns for discrete data, the current paper studies generalized Bayesian inference based on the Total Variation Distance (\TVD).
Letting $p(y,x)$, $p^{y|x}(y|x)$ and $p^x(x)$ denote the distributions of the true joint, conditional and marginal data generating mechanism and $p_{\theta}(x,y) = f_{\theta}(y|x)p^x(x)$, this means that we want to produce high posterior density in the regions of $\Theta$ where
\begin{IEEEeqnarray}{rCl}
    \TVD(p, p_{\theta})
    & = &
    \int_{\mathcal{X}}\sum_{y \in \mathcal{Y}}\left|p^{y|x}(y|x) - f_{\theta}(y|x)\right|dp^x(x)
    \nonumber \\
    & = &
    \mathbb{E}_{p^x}\left[ 
        \TVD(p^{y|x}(\cdot|x), f_{\theta}(\cdot|x))
    \right]
    \nonumber
\end{IEEEeqnarray}
takes relatively small values. 

\subsection{Why the \TVD?}\label{sec:why}

While the list of potential robust discrepancy measures is long, the \TVD is perhaps uniquely appealing.
First, the \TVD's very definition shows that it will seek values of $\theta$ producing well-calibrated probability models:
As the \TVD is the average absolute difference between the candidate probability model $f_{\theta}$ and the true data generating distribution, it assigns the highest posterior density to values of $\theta$ that best describe how $p$ allocates its probability mass.
This feature is not only intuitively attractive, but particularly suitable for choosing amongst probability measures on discrete spaces.
Second, the \TVD does not depend on any additional hyperparameters. Accordingly, we require no cross-validation or tuning strategy before inference is performed.
This is in stark opposition to a majority of alternatives: $\alpha$-, $\beta$- and $\gamma$-divergences are named after their hyperparameters while Minimum Stein Discrepancies \citep{MinimumSteinEstimator} and Maximium Mean Discrepancies \citep{FXMMD} depend on the choice of a kernel.
Third, and as we shall demonstrate next, the \TVD has universal robustness guarantees that are far stronger than those of most alternatives.
%
%
%

\subsection{Robustness}

Suppose the data-generating process is given by 
\begin{IEEEeqnarray}{rCl}
    c(y,x) = (1-\varepsilon)\cdot f_{\underline{\theta}}(y|x)p^x(x) + \varepsilon \cdot q(y,x),
    \label{eq:contamination_model}
\end{IEEEeqnarray}
where $\varepsilon \in (0,1)$ is the size of the contamination given by the distribution $q$, $f_{\underline{\theta}} \in \{f_{\theta}: \theta \in \Theta\}$ adequately describes the remaining data, and $p^x$ is the uncontaminated marginal distribution on $\mathcal{X}$.
Unlike with other discrepancy measures, any adverse effect $q$ has on inferring $\theta$ via the \TVD is bounded.
\begin{proposition}
For $p_{\theta}(y,x) = f_{\theta}(y|x)p^x(x)$, 
\begin{IEEEeqnarray}{rCl}
    \left|\TVD(c, p_{\theta}) - \TVD(p_{\underline{\theta}}, p_{\theta} )\right| \leq 2\varepsilon.
    \nonumber
\end{IEEEeqnarray}
     \label{proposition:TVD_robustness_idealized}
\end{proposition}

This result makes intuitive sense if one recalls that the \TVD selects for values of $\theta$ that correctly match the probability mass of the data-generating distribution.
Since the probability mass of contamination relative to the family $\{f_{\theta}:\theta \in \Theta\}$ is exactly $\varepsilon$, it logically follows that the \TVD should be off by a factor of order at most $\varepsilon$.
More striking still: The degree by which the contaminant $q$ is different from the unpolluted component has no impact, a property that makes the \TVD markedly different from the \KLD (see e.g. Figure \ref{fig:toy}).

\subsection{Estimating the \TVD}
These strong robustness properties make the \TVD a uniquely appealing discrepancy measure.
Unfortunately, we do not know the true data generating process $p$, which is needed to compute $\TVD(p, p_{\theta})$ exactly.
%
Accordingly, we instead need to estimate $\TVD(p, p_{\theta})$.
While this has inspired theoretically convincing prior work on estimating $\TVD(p, p_{\theta})$ \citep[e.g.][]{TVDestimate1985, TVDestimate2012}, the resulting estimators are typically both practically and computationally infeasible.
%
%
%
For instance, the estimator introduced by \cite{TVDestimate1985} not only requires the parameter space to be totally bounded, but also discretized.
%
%
%

To avoid practicality being an issue,
the current paper uses an estimator that is  theoretically inferior, but computationally superior  by orders of magnitude.
%
With $\phat_{n}(y,x)$, $\phat[y|x]_n(y|x)$ and $\widehat{p}^x_n(x)$  as before and for $\phat_{\theta, n}(y,x) = f_{\theta}(y|x)\phat[x]_n(x)$, we use 
\begin{IEEEeqnarray}{rCl}
    \TVD(\phat_n,  \phat_{\theta, n})
    & = &
    \mathbb{E}_{\phat[x]_n}\left[ 
        \TVD(\phat[y|x]_n(\cdot|x), f_{\theta}(\cdot|x))
    \right],
    \label{eq:TVD_loss}
\end{IEEEeqnarray}
%
which is similar to the \KLD estimator in \eqref{eq:ML=KLD}.
Throughout, the \textit{true} domain of the observations $y_i$ is given by $\mathcal{Y}_{\ast}$, which may be a subset of the \textit{model's} domain $\mathcal{Y}$.
This is natural under model misspecification:
For example, one may fit a Poisson regression model ($\mathcal{Y} = \mathbb{N}$) to the number of rainfall days in a year ($\mathcal{Y}_{\ast} = \{1,\dots,365\}$) for $n$ years.
%


\section{A RELIABLE ESTIMATE?}
\label{sec:estimator_properties}

While the \TVD has many desireable theoretical properties, it does have one decisive drawback relative to the \KLD: 
%
While the \KLD loss of \eqref{eq:ML=KLD} is \textit{linear} in the log likelihood functions, the \TVD loss of \eqref{eq:TVD_loss} is non-linear.
An immediate consequence is that performance guarantees of the \TVD estimator---such as convergence properties or finite-sample bounds---are much harder to derive.
%
%

The results of this section show that in spite of these complications, our estimator is generally very reliable.
A minor complication arises when the covariates $\{x_i\}_{i=1}^n$ are continuously-valued: In this case, we cannot study the estimator directly. Instead, we derive results for a surrogate estimator that relies on kernel density estimation, but can be made arbitrarily close to our naive estimator.

We find that under mild conditions, our estimator for the \TVD satisfies exponential concentration inequalities. This immediately allows us to conclude that (i) it converges to its target almost surely and that (ii) the robustness property of Proposition \ref{proposition:TVD_robustness_idealized} applies to the estimated objects, too.
With additional labour, one can also show that the minimizers of $\TVD(\phat_n, \phat_{\theta, n})$ are strongly consistent for the minimizer of $\TVD(p, p_{\theta})$.
Results with proof sketches are derived in full detail in the Appendix.

\subsection{Exponential Concentration Inequalities}
As the arguments and rates are slightly different, we give separate results for discrete-valued and continuous-valued covariates.

\begin{proposition}
    If $\{x_i, y_i\}_{i=1}^n$ are both discrete-valued and sampled i.i.d. from a probability distribution $\mathbb{P}$ such that $y_i \in \mathcal{Y}$, $x_i \in \mathcal{X}$ and $|\mathcal{Y}| = K_y$, $|\mathcal{X}| = K_x$ for some $K_x, K_y \in \mathbb{N}$, then it holds that pointwise for any $\theta \in \Theta$ and for any $\varepsilon > 0$ and with probability at least $1-\delta_n$,
    \begin{IEEEeqnarray}{rCl}
        \left|
            {\TVD}(\phat_n, \phat_{\theta,n})
            -
            \TVD(p, p_{\theta})
        \right| < \varepsilon
        \nonumber
    \end{IEEEeqnarray}
    where $\delta_n = (2^{K_y+K_x+1} - 2^2)e^{-n\varepsilon^2/2}$.
    \label{proposition:concentration_discrete}
\end{proposition}
\begin{proofsketch}
    One can use the triangle inequality and the fact that $|x + y| \geq ||x| - |y||$ to show that 
    \begin{IEEEeqnarray}{rCl}
    &&
\left|
            {\TVD}(\phat_n, \phat_{\theta,n})
            -
            \TVD(p, p_{\theta})
        \right|
        \nonumber \\
     & \leq & 
     \TVD(\phat_n, p) + \TVD(p_{\theta}, \phat_{\theta, n}).
     \nonumber
    \end{IEEEeqnarray}
    For either of these two expressions, exponential concentration inequalities apply \citep[e.g.][]{concentrationTVDDiscrete_older} so that one can use a union bound to show that the result holds. 
\end{proofsketch}

Similar arguments can be used for continuous covariates if one studies the surrogate estimator
\begin{IEEEeqnarray}{rCl}
    \TVD(\phat[h_n]_n, \phat_{\theta,n,h_n})
    & \approx &
    \TVD(\phat_n, \phat_{\theta, n})
    \nonumber
\end{IEEEeqnarray}
where for a suitable kernel $K$ and bandwidth $h_n$, the kernel density smoothed estimator $\phat[h_n]_n$ of $p$ is 
\begin{IEEEeqnarray}{rCl}
         \phat[h_n]_n(x,y) & = & 
         \frac{1}{n}\sum_{i=1}^n \delta_{y_i}(y) \cdot \frac{1}{h_n^d}K\left(\frac{x_i - x}{h_n}\right).
         \nonumber
\end{IEEEeqnarray}

\begin{proposition}
    Suppose $\{y_i\}_{i=1}^n$ are discrete with $|\mathcal{Y}|  = K_y$ while $\{x_i\}_{i=1}^n$ are continuous with $\mathcal{X} \subseteq \mathbb{R}^d$ and that $\{x_i, y_i\}_{i=1}^n$ are sampled i.i.d from a probability distribution $\mathbb{P}$.
    Assume that the true marginal distribution $\mathbb{P}_x$ of $x$ admits a density $p^x$ that is  absolutely continuous with respect to the Lebesgue measure.
    Further, suppose that $\int_{\mathcal{X}^2}K(x,x')dxdx' = 1$ and $h_n \to 0$ while $n\cdot h_n \to \infty$ as $n\to \infty$.  
    Then it holds for $n$ large enough, any $\theta \in \Theta$, any $\varepsilon > 0$ and with probability at least $1-\delta_n$,
    \begin{IEEEeqnarray}{rCl}
        \left|
            {\TVD}(\phat[h_n]_n, \phat_{\theta,n,h_n})
            -
            \TVD(p, p_{\theta})
        \right| < \varepsilon
        \nonumber
    \end{IEEEeqnarray}
    where $\phat_{\theta,n,h_n}(y,x) = p_{\theta}(y|x)\phat[x,h_n]_n(x)$.
    Here, for a constant $r$ depending only on $K_y$, and for $n_y = \sum_{i=1}^n\delta_y(y_i)$ denoting the number of samples for which $y_i = y$,  we have $\delta_n = \exp\{-\min_{y \in \mathcal{Y}}n_{y} \cdot \varepsilon^2\cdot r\}$.
    \label{proposition:concentration_continuous}
\end{proposition}
\begin{proofsketch}
    The proof proceeds along similar lines as in the discrete case, but is complicated by the fact that for $q(x,y) = \phat[x|y, h_n]_n(x|y)p^{y}(y)$, one additionally upper bounds
     \begin{IEEEeqnarray}{rCl}
         \TVD(\phat[h_n]_n, p) 
         &  \leq  &
        \TVD(\phat[h_n]_n, q)
        +
        \TVD(q, p),
        \nonumber 
    \end{IEEEeqnarray}
    which in turn we can further upper bound by
    \begin{IEEEeqnarray}{rCl}
        & \leq &
        \TVD(\phat[y, h_n]_n, p^y)
        +
        \sup_{y \in \mathcal{Y}} \TVD(\phat[x|y, h_n]_n(\cdot|y), p^{x|y}(\cdot|y)).
        \nonumber
     \end{IEEEeqnarray}
    The first term vanishes exponentially fast by the same arguments deployed for the discrete case. The second term requires a conditionalization argument together with an upper bound which introduces $\min_{y \in \mathcal{Y}}n_y$ into the bound. 
    Lastly, one uses a union bound argument to put everything together.
\end{proofsketch}
%

%
While the last result does not apply to the actual estimator of interest in \eqref{eq:TVD_loss}, it does apply to its kernel-density based surrogate.
In spite of this, all experiments in the current paper use \eqref{eq:TVD_loss} rather than its surrogate.
Why do we insist on using an estimate of likely inferior theoretical quality?

The answer is threefold: 
Firstly, though a \TVD estimate based on kernel density estimation is theoretically appealing, it would add two orders of magnitude to the computational complexity of our algorithm.
This renders the kernel density surrogate practically infeasible for most situations, a feature present in pre-existing estimators for the \TVD \cite[e.g.][]{TVDestimate1985, TVDestimate2012}.
Secondly, it is reasonable to expect  the behaviour of the naive estimates to be fairly similar to that based on kernel density estimates.
In fact, one can make $\phat[x,h_n]_n$ arbitrarily close to $\phat[x]_n$ for small enough $h_n$. 
For instance, Proposition \ref{proposition:concentration_continuous} holds for a uniform kernel and $h_n = \eta \cdot n^{-1/5}$, which (even if $n=1$) can be made arbitrarily close to the empirical measure for $\eta \to 0$.
%
Thirdly and most persuasively, our empirical results demonstrate that the naive estimator of \eqref{eq:TVD_loss} performs convincingly, even if $\mathcal{X}\subseteq\mathbb{R}^d$.

%

%

\subsection{Almost sure convergence}

The exponential concentration inequalities derived in Propositions \ref{proposition:concentration_discrete} and \ref{proposition:concentration_continuous} are attractive for many reasons.
Most importantly, they provide computable finite-sample guarantees.
Further, they also imply almost sure convergence by the Borel-Cantelli Lemma.
%

\begin{corollary}
    Under the conditions of Proposition \ref{proposition:concentration_discrete}, 
    ${\TVD}(\phat_n, \phat_{\theta,n})
           \overset{a.s.}{\longrightarrow}
           \TVD(p, p_{\theta})$ 
    as $n\to \infty$.
    Under the conditions of Proposition \ref{proposition:concentration_continuous}, 
    ${\TVD}(\phat_n, \phat_{\theta,n,h_n})
            \overset{a.s.}{\longrightarrow}
            \TVD(p, p_{\theta})$ 
    as $\min_{y\in\mathcal{Y}} n_y\to \infty$.
    \label{corollary:as_convergence}
\end{corollary}

\subsection{Provable robustness}

Proposition \ref{proposition:TVD_robustness_idealized} demonstrated that the \textit{true} \TVD is robust. Further, Propositions \ref{proposition:concentration_discrete} and \ref{proposition:concentration_continuous} showed that the estimated \TVD rapidly converges to the truth.
Intuitively then, it stands to reason that the \textit{estimated} \TVD is also robust.
The following result confirms this.

\begin{corollary}
Pick any $\eta > 0$.
    Under the conditions of Proposition \ref{proposition:concentration_discrete} and with $\phat_n$ an empirical measure constructed from $c$ as in \eqref{eq:contamination_model}, there is $N$ such that for all $n \geq N$, 
    \begin{IEEEeqnarray}{rCl}
        \left|
            {\TVD}(\phat_n, \phat_{\theta,n})
            -
            \TVD(p_{\underline{\theta}}, p_{\theta})
        \right|
        \leq 2\varepsilon + \eta
        \nonumber
    \end{IEEEeqnarray}
    holds with probability one.
    Similarly, under the conditions of Proposition \ref{proposition:concentration_continuous}, it holds that there exists $N$ such that for all $n$ for which $\min_{y \in \mathcal{Y}}n_y \geq N$, 
    \begin{IEEEeqnarray}{rCl}
        \left|
            {\TVD}(\phat[h_n]_n, \phat_{\theta,n})
            -
            \TVD(p_{\underline{\theta}}, p_{\theta})
        \right|
        \leq 2\varepsilon + \eta
        \nonumber
    \end{IEEEeqnarray}
    holds with probability one.
    \label{corollary:robustness_estimator}
\end{corollary}

\subsection{Consistency}

In spite of being firmly rooted within the Bayesian paradigm, the inference method we will present in the next section relies on computing a perturbed form of the minimizers $\theta_n = \argmin_{\theta \in \Theta}\TVD(\phat_n, \phat_{\theta, n})$.
Thus, we are interested in the convergence properties of $\theta_n$.
While the convergence properties of similar estimators have been studied before \citep{TVDestimate1985}, the analysis we employ is drastically different. This is because unlike the estimator of \citet{TVDestimate1985}, we do not require $\Theta$ to be discretized into a grid.

In spite of this, we can show strong consistency of $\theta_n$ with respect to $\theta^\ast = \argmin_{\theta \in \Theta} \TVD(p, p_{\theta})$ under
mild differentiability conditions (See Assumption 1, Appendix \ref{app:consistency}). 
%
%
%
\begin{proposition}
Suppose Assumption 1 holds.
    If $\theta_n = \argmin_{\theta \in \Theta}\TVD(\phat_n, \phat_{\theta, n})$
    is almost surely unique for all $n$ large enough and the conditions of Proposition \ref{proposition:concentration_discrete} hold, $\theta_n \overset{a.s.}{\longrightarrow} \theta^\ast$ as $n \to \infty$.
    Similarly, if 
    $\theta_n = \argmin_{\theta \in \Theta}\TVD(\phat[h_n]_n, \phat_{\theta, n})$
    is almost surely unique for all $n$ large enough and the conditions of Proposition \ref{proposition:concentration_continuous} hold, $\theta_n \overset{a.s.}{\longrightarrow} \theta^\ast$ as  $\min_{y\in\mathcal{Y}}n_y \to \infty$.
    \label{proposition:consistency}
\end{proposition}
\begin{proofsketch}
We give the proof for the discrete case only, as the continuous case follows similar arguments.
    First, we show that 
    \begin{IEEEeqnarray}{rCl}
    0 \leq 
        \TVD(p, p_{\theta_n}) - \TVD(p, p_{\theta^\ast})
        \leq 
         2\cdot\TVD(p_n, p_{\ast}).
        \nonumber 
    \end{IEEEeqnarray}
    Further, $\TVD(\phat_n, p)$ goes to zero almost surely as $n\to\infty$.
    Since we care about the limiting value of $\theta_n$, we may confine the analysis to large enough $n$ (say $n>N$) for which $2\cdot\TVD(\phat_n, p) < \varepsilon$. 
    Because  $\TVD(p, p_{\theta^\ast}) - \TVD(p, p_{\theta_n})< \varepsilon$ implies that $\theta_n \in B_{\varepsilon}$, restricting attention to $n>N$ is equivalent to restricting attention to the compact set $B_{\varepsilon}$.
    Thanks to this and the finite gradients,
    $\TVD(p, p_{\theta^\ast}) - \TVD(p, p_{\theta_n})$   converges to zero uniformly over $B_{\varepsilon}$.
    Together with the fact that $\theta_n \in B_{\varepsilon}$ for $n>N$, this implies the result.
\end{proofsketch}

\section{NONPARAMETRIC LEARNING}
\label{sec:nonparametric_learning}

In the context of misspecification, choosing a prior over the parameter space $\Theta$ is conceptually unappealing almost by definition:
On the one hand, we readily admit that the probability model $f_{\theta}$ is misspecified, meaning that the interpretation of `good' values for $\theta$ is not straightforward. 
On the other hand, we are forced into having our quantification of uncertainty depend on a prior belief over the parameter $\theta$.
As a consequence, the inferred uncertainties are not straightforwardly interpretable in a Bayesian sense.

To avoid these complications, we opt for a different strategy: Instead of choosing a prior over $\theta$, we impose an uninformative prior \textit{directly} on the data-generating mechanism.
This approach has recently been advocated in a series of papers \citep{SimonNPL, EdwinNPL} under the name `Bayesian Nonparametric Learning' (\NPL) and has three main benefits: Firstly, it is suitable for generalized Bayesian inference with arbitrary loss functions. Secondly, it perfectly suits the misspecified setting. Thirdly, it produces exact inferences at low computational cost. For completeness, we give a brief description of the core components and the corresponding inference algorithm.

\subsection{Bayesian Nonparametric Learning}
Suppose that we have access to the true data-generating mechanism $p$ rather than some sample $\{x_i, y_i\}_{i=1}^n$.
In this case, we have \textit{no need} for uncertainty. 
In fact, we could simply compute
\begin{IEEEeqnarray}{rCl}
    \theta^\ast = \theta(p) & = &
    \argmin_{\theta \in \Theta}{\TVD}(p, p_{\theta}).
    \nonumber
\end{IEEEeqnarray}
In practice of course, this is impossible. In fact, it is this very impossibility that necessitates both the Frequentist and Bayesian notions of uncertainty.
Taking this observation to heart, \citet{SimonNPL} and \citet{EdwinNPL} have asked: Instead of quantifying uncertainty about `good' parameter values by placing a prior on $\theta$, why not place a prior on $p$?
In its characterization of the desiderata for Bayesian uncertainty, this approach closely tracks the classical bootstrap \citep{OriginalBootstrap} and its Bayesian counterpart \citep{BayesBootstrap}.

While placing a prior over $p$ is easier said than done, \citet{SimonNPL} and \citet{EdwinNPL} show that Dirichlet Processes (\DP{}s) are a computationally efficient way of doing this.
For a measure $p_{\pi}$ over $\mathcal{X} \times \mathcal{Y}$ enconding our prior beliefs about $p$ and a
scalar $\alpha$ determining the strength of this belief, we follow the approach in \citet{EdwinNPL} and define the following (nonparametric) Bayesian prior about the data-generating mechanism:
\begin{IEEEeqnarray}{rCl}
    p & \sim &
    \DP(\alpha, p_{\pi}).
    \nonumber
\end{IEEEeqnarray}
Because we have assumed that our observations have been generated independently and identically distributed, this choice of prior yields a conjugate closed-form \DP posterior as 
\begin{IEEEeqnarray}{CCl}
    p| \{x_i, y_i\}_{i=1}^n  \sim 
    \DP(\alpha + n, p_{\pi,n});
    \nonumber \\
    p_{\pi,n} =
    \frac{\alpha}{\alpha + n}\cdot p_{\pi}
    +
    \frac{1}{\alpha + n}\cdot \sum_{i=1}^n\delta_{(x_i, y_i)}.
    \label{eq:Edwins_posterior}
\end{IEEEeqnarray}
This suggests sampling $\{p^{(i)}\}_{i=1}^B$ from $\DP(\alpha + n, p_{\pi,n})$ and then computing a sample $\{\theta_i\}_{i=1}^B$, where $\theta_i = \theta(p^{(i)})$.
To make sampling from a \DP computationally feasible, some form of truncation limit $T$ is required.
Again, we follow the suggestions of \citet{EdwinNPL}, yielding the \textit{Posterior Boostrap Sampling} algorithm.
%

%
\begin{algorithm}[h!]
	\caption{\text{Posterior Bootstrap Sampling}}
   \label{Algorithm_BOCPDMS}
\begin{algorithmic}[0]
   \State {\bfseries Input:} $\{x_i, y_i\}_{i=1}^n$, $\alpha$, $p_{\pi}$, $T$
   \For{j = 1, 2, \dots B}
    \State draw pseudo-samples $(\widetilde{y}^{(j)}_i, \widetilde{x}^{(j)}_i)_{1:T}  \overset{i.i.d.}{\sim} p_{\pi}$
    \State draw $(w^{(j)}_{1:n}, \widetilde{w}^{(j)}_{1:T}) \sim \text{Dir}(1,\dots,1, \alpha/T, \dots, \alpha/T)$
    \State get $p^{(j)} = \sum_{i=1}^nw^{(j)}_i \delta_{(y_i, x_i)} + \sum_{k=1}^T \delta_{(\widetilde{y}^{(j)}_i, \widetilde{x}^{(j)}_i)}$
    \State get $p^{(j)}_{\theta, n}(x,y) = p_{\theta}(y|x)\left[\sum_{y \in \mathcal{Y}}p^{(j)}(x,y) \right]$
    %
    \State Compute $\theta^{(j)} = \argmin_{\theta \in \Theta}{\TVD}(p^{(j)}, p^{(j)}_{\theta, n})$
   \EndFor
    \State \Return{posterior bootstrap sample $\theta^{(1:B)}$}
\end{algorithmic}
\end{algorithm}

This Bayesian inference scheme has three rare and desirable properties: It is simple, embarassingly parallel, and produces independent parameter samples.

\subsection{Practical considerations: $\alpha$ and $p_{\pi}$}

There are two main knobs for tuning the above algorithm: The prior $p_{\pi}$ and the scalar $\alpha$.
Throughout our experiments, we use the limiting case of $\alpha \to 0$, which automatically eliminates the need to specify $p_{\pi}$ (see \eqref{eq:Edwins_posterior}). As pointed out by \citet{EdwinNPL}, this has the interpretation of positing a maximally uninformative prior belief about $p$.
Computationally, the algorithm is equivalent to a generalized form of the Bayesian Bootstrap \citep{BayesBootstrap} introduced by \citet{LyddonBayesBootstrap}.

This choice of $\alpha$ is thus  justified from a conceptual as well as a practical standpoint.
Conceptually, it reflects the fact that we have no clear idea about the nature of $p$---since if we had, we could specify a model family $\{f_{\theta}:\theta \in \Theta\}$ that is not drastically misspecified.
On a practical level, it simplifies the inference algorithm by eliminating two hyperparameters ($\alpha$ and $p_{\pi}$) and leads to a considerable speedups.
%
%

\begin{figure*}[th!]
    \centering
    \includegraphics[trim= {0.0cm 0.4cm 0.cm 0.0cm}, clip,width=\textwidth]{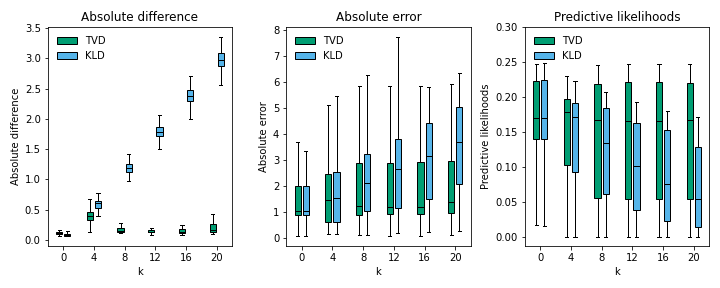}
    \caption{
    Difference in inference outcomes for the $\varepsilon$-contamination model of \eqref{eq:eps_contamination_experiment} between using the \textcolor{TVDColor}{\textbf{TVD}} and the \textcolor{KLDColor}{\textbf{KLD}} as $k$ is varied and $\varepsilon=0.15$.
    \textbf{Left}: Absolute difference between inferred and true value of $\lambda$;
    \textbf{Middle}: Absolute out-of-sample prediction error; 
    \textbf{Right}: Predictive likelihood on out-of-sample data. 
    }
    \label{fig:eps}
\end{figure*}

\subsection{Finding the minima $\theta^{(j)}$}

A more implicit knob in tuning the algorithm is the sub-routine one chooses to find the minimizer $\theta^{(j)}$.
Minimizing ${\TVD}(p^{(j)}, p^{(j)}_{\theta, n})$ is generally very difficult: The function will not be convex in $\theta$ everywhere. Worse still, it will in fact be equal to its upper bound for most values of $\theta$. 
This makes it crucial to find good initial values from which to start a gradient-based optimization process. 
To address this, we compute the minimizers by using  maximum likelihood estimates as initializers and then compute the (possibly local) minima of ${\TVD}(p^{(j)}, p^{(j)}_{\theta, n})$ using the Broyden–Fletcher–Goldfarb–Shanno  (BFGS) algorithm.

\section{EXPERIMENTS}
\label{sec:experiments}
We verify the performance of our method on a number of synthetic and real world data examples.
First, we study two canonical synthetic data examples: An $\varepsilon$-contaminated Poisson model and a zero-inflated binomial model.
We find that our method recovers parameter estimates that are close to those of the uncontaminated data generating process and improve out-of-sample predictive performance. 
%
%
Next, we investigate performance on real world data with two standard classification models: Probit regressions and Neural Networks.
For both models and across all datasets, we find that our method improves predictive likelihoods.
Additional experimental details can be found in the Appendix. All code is publicly available at \texttt{https://github.com/CENSORED}.

\begin{figure*}[ht!]
    \centering
    \includegraphics[trim= {0.0cm 0.35cm 0.cm 0.0cm}, clip,width=\textwidth]{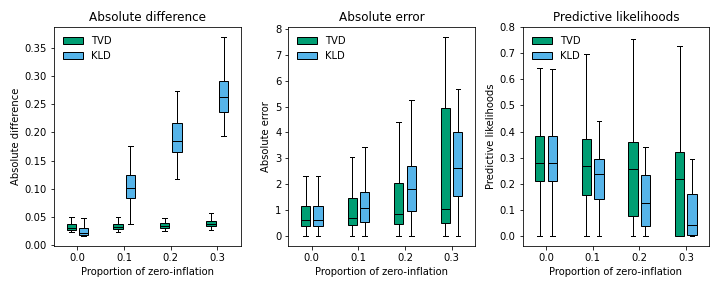}
    \caption{
    Difference in inference outcomes for the zero-inflation model between using the \textcolor{TVDColor}{\textbf{TVD}} and the \textcolor{KLDColor}{\textbf{KLD}} as the proportion $\varepsilon$ of zeros is varied.
    \textbf{Left}: Absolute difference between inferred and true value of $\beta$;
    \textbf{Middle}: Absolute out-of-sample prediction error; 
    \textbf{Right}: Predictive likelihood on out-of-sample data. 
    }
    \label{fig:zero}
\end{figure*}

\subsection{$\varepsilon$-contamination}
We consider an $\varepsilon$-contaminated Poisson model based on \eqref{eq:contamination_model}: A proportion $(1 - \varepsilon)$ of the data come from a standard Poisson model while $\varepsilon \in (0,1)$ of the data come from a contamination distribution.
Throughout the experiments, we fix the mean of the Poisson distribution as $\lambda = 3$, the proportion $\varepsilon=0.15$ and consider contamination in form of an offset by a constant $k$.
%
In other words, we generate our data as
\begin{IEEEeqnarray}{rCl}
    y_i & \overset{iid}{\sim} & \text{Poisson}(\lambda) + \text{Bernoulli}(\varepsilon) \cdot k
    \label{eq:eps_contamination_experiment}
\end{IEEEeqnarray}
For a fixed value of $k$, we generate 100 datasets of 500 observations. 
Each of these is split into train and test datasets of size $n_\text{train} = 400, n_\text{test} = 100$. 
We then use the inference method sketched in Algorithm \ref{Algorithm_BOCPDMS} to infer two misspecified Poisson models by drawing $B=1000$ samples, 
one based on minimizing the Kullback-Leibler divergence (\KLD) and one based on minimizing the \TVD. 
We also perform standard Bayesian inference using Stan \citep{stan}, but the results are not competitive and deferred to the Appendix \ref{apx:experiments}. 

Figure \ref{fig:eps} reports the results with a focus on three key quantities: 1.\ the absolute difference between the true and inferred value of $\lambda$, 2.\ the absolute error on the test data and 3.\ the predictive likelihood. 
The exact computation underlying these quantities can be found in Appendix \ref{app:evaluation_criteria}.

As shown in the left panel of \figref{fig:eps}, minimizing the \TVD produces parameters similar to those of the \KLD in the absence of contamination ($k = 0$) and produces consistently better estimates as the degree of contamination becomes more severe.
Similarly, the \TVD improves out-of-sample prediction. Notably, its median absolute error remains essentially constant. Lastly, Figure \ref{fig:eps} also shows that the \TVD consistently improves predictive likelihoods relative to the \KLD---even under increasingly extreme contamination. This improvement in calibration is perhaps not surprising, as it is predicted by the theory outlined in \secref{sec:why}.

\subsection{Zero-inflation}

We also consider a zero-inflated binomial regression model where the probability of success $\pi$ out of $m=8$ trials is modeled by a categorical covariate.
Casting this in the language of \eqref{eq:contamination_model}, this means that the contaminating distribution is a dirac delta at zero.
%
This implies the following model:
\begin{IEEEeqnarray*}{rCl}
    y_i|x_i & \overset{iid}{\sim} &  \text{Binomial}(8, \pi_i) \cdot (1- \text{Bernoulli}(\varepsilon))
    \nonumber \\
    \pi_i &=& \text{logit}^{-1}(0.8 + 0.25 x_i)\\
    x_i &\overset{iid}{\sim} & \text{Categorical}( \mathbf{p}, 4), \quad p_1 = \ldots = p_4 = 1/4.
\end{IEEEeqnarray*}
We set $n=1000$, $n_\text{train} = 800$, and $n_\text{test} = 200$. 
We are interested in the same three quantities as in the $\varepsilon$-contamination example and report these in Figure \ref{fig:zero}.
Instead of $\lambda$, we now report the absolute difference between the inferred and true value of $\beta = 0.25$.
As before, the \TVD guarantees parameter estimates closer to the uncontaminated data-generating process 
than those of the \KLD. 
In turn, this yields superior out-of-sample prediction and predictive likelihoods.

\begin{figure}[hb!]
\centering
\includegraphics[trim= {-0.5cm 9.25cm 0.cm 0.0cm}, clip, width=0.5\textwidth]{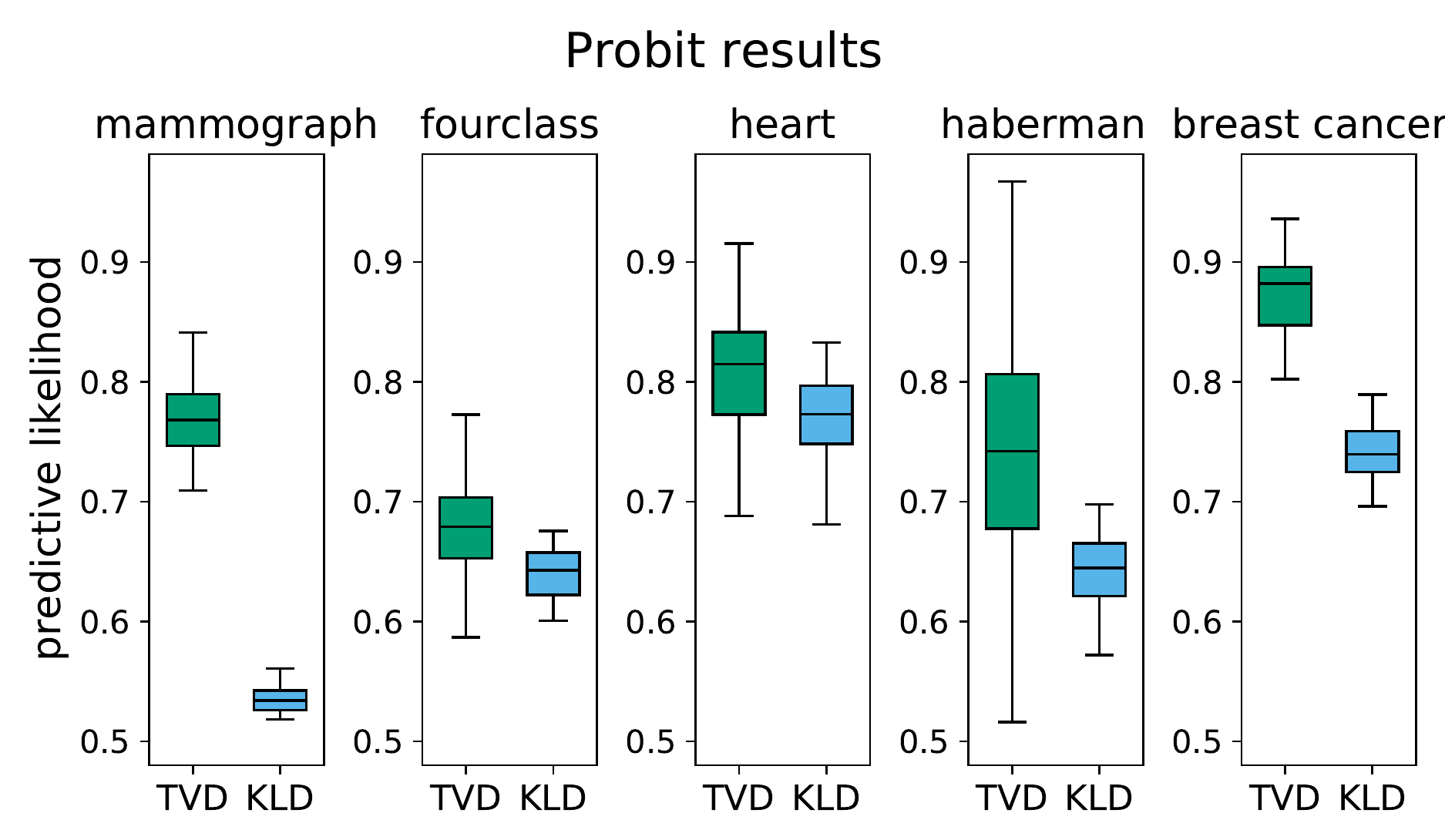}
\hspace*{0.25cm}
\includegraphics[trim= {0.2cm 9.25cm 0.cm 0.0cm}, clip, width=0.5\textwidth]{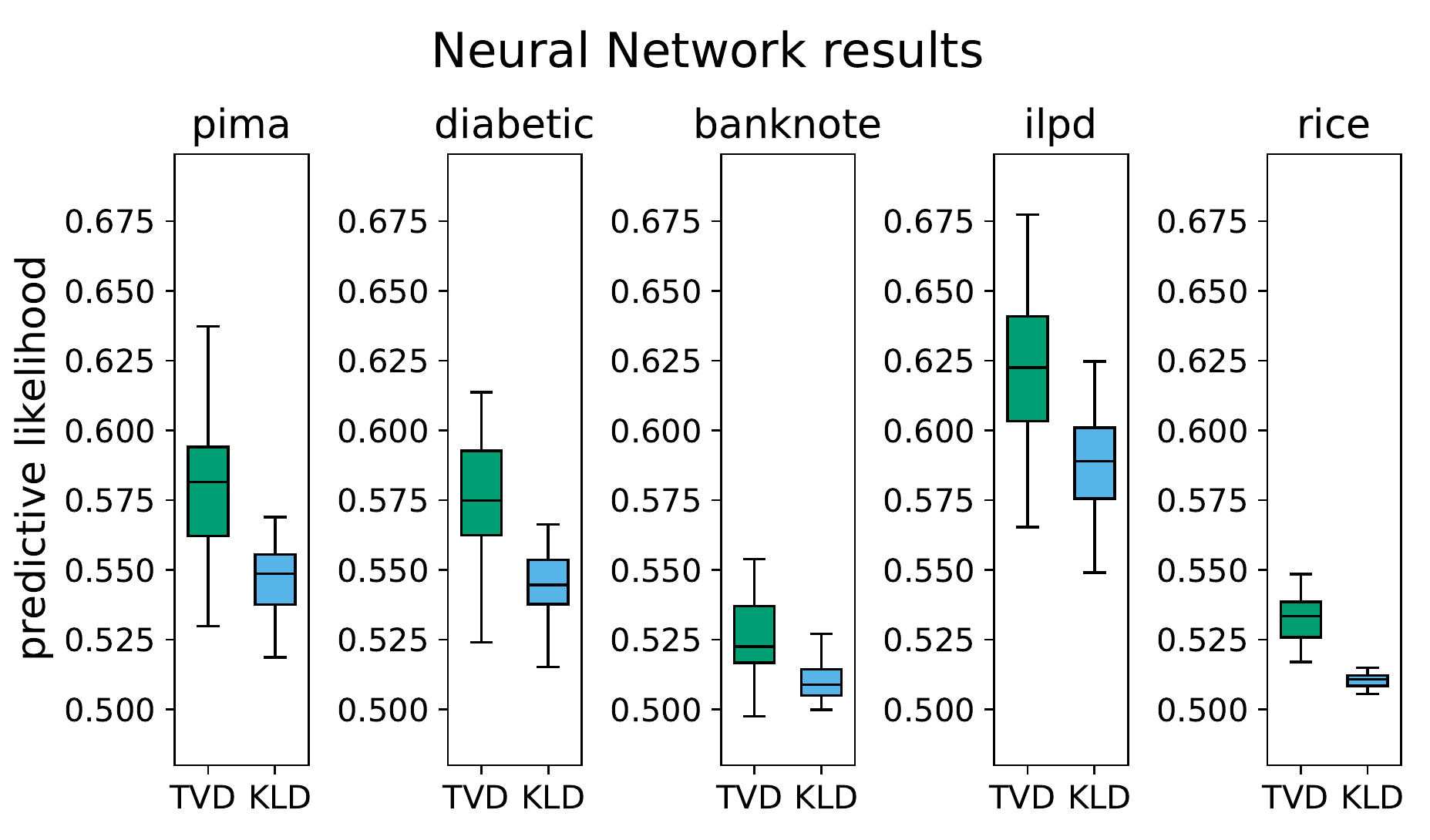}
\caption{Predictive likelihoods for the Probit model (top) and a single-layer Neural Networks (bottom). 
}
\label{fig:probit_NN_results}
\end{figure}

\subsection{Real world datasets}
Classification is another important problem setting with discrete-valued outcomes.
To this end, we evaluate our approach on two different classification models on a range of real world datasets taken from the UCI repository \citep{UCI}. 
For each dataset, we produce $50$ random training:test (90\%:10\%) splits of the data and compare the inference outcomes of Algorithm \ref{Algorithm_BOCPDMS} with $B=1000$ samples based on our \TVD-minimizing approach against those obtained by minimizing the \KLD. 

\subsubsection{Probit}
The Probit model is a canonical statistical model for binary classification. For our evaluation, we select $5$ datasets from the UCI repository with discrete covariates.
None of these datasets are easy to describe with simple off-the-shelf models which strongly suggests the presence of model misspecification.
Our findings confirm this: Using the robust \TVD produces a model with better predictive calibration than the \KLD on all five datasets (see top row of \figref{fig:probit_NN_results}).


\subsubsection{Neural Network}
Neural networks are another important Machine Learning classification model. Again, we select five datasets from UCI with binary outcomes. Unlike the datasets in the Probit section, these datasets have continuous covariates.
In spite of this added complication, the bottom row of Figure \ref{fig:probit_NN_results} shows that we find similar improvements in the predictive calibration as for the Probit models.
All Neural Network examples were run with a single hidden layer of $50$ nodes. 

\section{CONCLUSION}
We propose a new generalized Bayesian inference method using the Total Variation Distance (\TVD) to robustify the inference of models for discrete-valued outcomes. 
The resulting inference procedure is based on an estimator of the \TVD which possesses a range of desirable theoretical properties. 
In practice, our method yields significantly improved inference outcomes under model misspecification. 
Since we found no significant loss of efficiency in the absence of misspecification, we conclude that our method is a useful generic robustness approach.

\subsection*{Acknowledgements}

JK is funded by the EPSRC grant EP/L016710/1 as part of the Oxford-Warwick  Statistics Programme (OxWaSP). JK is additionally funded by the Facebook Fellowship Programme and the London Air Quality project at the Alan Turing Institute for Data Science and AI as part of the Lloyd's Register Foundation programme on Data Centric Engineering. This work was furthermore supported by The Alan Turing Institute for Data Science and AI under EPSRC grant EP/N510129/1 in collaboration with the Greater London Authority.
Furthermore, LV is supported by the Leverhulme Trust.

\bibliography{library}

\appendix

\section{Proofs}

We give a detailed account for the concentration inequalities as well as related results derived in the main paper.

\subsection{Proposition \ref{proposition:TVD_robustness_idealized}}

\begin{proof}
One simply observes that by definition,
\begin{IEEEeqnarray}{rCl}
    && \left|\TVD(c, p_{\theta}) - \TVD(p_{\underline{\theta}}, p_{\theta})\right|
    \nonumber \\
    & = &  
    \left|
        (1-\varepsilon)\TVD(p_{\underline{\theta}}, p_{\theta})
        + \varepsilon \TVD(q, p_{\theta}) - \TVD(p_{\underline{\theta}}, p_{\theta})
    \right|
    \nonumber \\
    & \leq & 
    \left|
        \varepsilon\TVD(q, p_{\theta})
        + \varepsilon \TVD(p_{\underline{\theta}}, p_{\theta})
    \right|
    \leq
    2\varepsilon,
    \nonumber
\end{IEEEeqnarray}
which completes the proof.
\end{proof}

\subsection{Corollary \ref{corollary:as_convergence}}

\begin{proof}
    We only give proof for the discrete case, as the arguments are the same for the continuous case. Denoting
    $A_n = \{\left|
            {\TVD}(\phat_n, \phat_{\theta,n})
            -
            \TVD(p, p_{\theta})
        \right| > \varepsilon\}$,
     \begin{IEEEeqnarray}{rCl}
         \sum_{i=1}^{\infty}
         \mathbb{P}(A_n)
         & \leq &
         (2^{K_y+K_x+1} - 2^2)
         \sum_{i=1}^n e^{-n\varepsilon^2/2} < \infty.
         \nonumber
     \end{IEEEeqnarray}
     This immediately implies that we can apply the Borel-Cantelli Lemma to conclude that
     \begin{IEEEeqnarray}{rCl}
         \mathbb{P}(\lim_{n\to\infty}A_n)
         \leq 
         \mathbb{P}(\lim\sup_{n\to\infty}A_n)
         = 0.
         \nonumber
     \end{IEEEeqnarray}
     As $\varepsilon$ was chosen arbitrarily, the proof is complete.
\end{proof}

\subsection{Corollary \ref{corollary:robustness_estimator}}

\begin{proof}
    We only give proof for the discrete case, as the arguments are the same for the continuous case.
    By Corollary \ref{corollary:as_convergence}, we know that there exists $N$ such that with probability one,
    $ \left|
            {\TVD}(\phat_n, \phat_{\theta,n})
            -
            \TVD(c, p_{\theta})
        \right|
        \leq \eta$ for all $n\geq N$.
    Thus, for $n\geq N$ it holds that
    \begin{IEEEeqnarray}{rCl}
        &&\left|
            {\TVD}(\phat_n, \phat_{\theta,n})
            -
            \TVD(p_{\underline{\theta}}, p_{\theta})
        \right|
        \nonumber \\
        & \leq &
        \left|
            {\TVD}(\phat_n, \phat_{\theta,n})
            -
            \TVD(c, p_{\theta})
        \right|
        \nonumber \\
        & & 
        +
        \left|
            {\TVD}(p_{\underline{\theta}}, p_{\theta})
            -
            \TVD(c, p_{\theta})
        \right|
        \nonumber \\
        & \leq &
        \eta + 2\varepsilon
        \nonumber,
    \end{IEEEeqnarray}
    which completes the proof.
\end{proof}

\subsection{Proposition \ref{proposition:concentration_discrete}}

\begin{proof}
    The proof proceeds in two steps. 
    First, we bound ${\TVD}(\phat_n, \phat_{\theta,n})$ from above and below using the same terms.
    Second, we show that only one of these terms (namely 
    $\mathbb{E}_{p^x}[\TVD(p^{y|x}, f_{\theta})]$) does not satisfy an exponential concentration towards zero.

    For simplicity, we first give the proof for countable spaces $\mathcal{X}$ and then extend it to uncountable Euclidean spaces $\mathcal{X}$ later.
    Writing $p_{\theta}(x,y) = p_{\theta}(x|y)p^x(x)$ as well as $\phat_{\theta, n}(x,y) = f_{\theta}(x|y)\phat[x]_{n}(x)$, we begin by noting that
    \begin{IEEEeqnarray}{rCl}
        \TVD(\phat_n, \phat_{\theta, n})
        & = &
        \frac{1}{2}\sum_{x \in \mathcal{X}}\sum_{y \in \mathcal{Y}}\left|
            \phat_n(x,y) - \phat_{\theta, n}(x,y)
        \right|.
        \nonumber
    \end{IEEEeqnarray}
    Defining further the functions $\Delta(x,y) = \phat_n(x,y) - p(x,y)$, $\Delta_{\theta, \ast}(x,y) = p(x,y) - p_{\theta}(x,y)$ and $\Delta_{\theta, n}(x,y) = p_{\theta}(x,y) - \phat_{\theta, n}(x,y)$, we can rewrite and lower bound the above as
     \begin{IEEEeqnarray}{rCl}
     &&
        \frac{1}{2}\sum_{x \in \mathcal{X}}\sum_{y \in \mathcal{Y}}\Big|
            \Delta(x,y) + \Delta_{\theta, \ast}(x,y) + \Delta_{\theta, n}(x,y)
        \Big|
        \nonumber \\
     & \geq &
        \frac{1}{2}\sum_{x \in \mathcal{X}}\sum_{y \in \mathcal{Y}}\Big| |\Delta_{\theta, \ast}(x,y)|
            - 
            |\Delta(x,y) + \Delta_{\theta, n}(x,y)|
        \Big|
        \nonumber \\
    & \geq &
        \frac{1}{2}\sum_{x \in \mathcal{X}}\sum_{y \in \mathcal{Y}}\Big| |\Delta_{\theta, \ast}(x,y)|
            - 
            |\Delta(x,y)|
            - 
            |\Delta_{\theta, n}(x,y)|
        \Big|
        \nonumber \\
    & \geq &
        \Bigg|\frac{1}{2}\sum_{x \in \mathcal{X}}\sum_{y \in \mathcal{Y}} |\Delta_{\theta, \ast}(x,y)|
            - 
            |\Delta(x,y)|
            - 
            |\Delta_{\theta, n}(x,y)|
        \Bigg|
        \nonumber \\
    & = &
        \big| \TVD(p, p_{\theta}) - \TVD(\phat_n, p) - \TVD(p, \phat_{\theta, n}) \big|.
        \nonumber \\
    & \geq &
        \TVD(p, \phat_{\theta}) - \left[\TVD(\phat_n, p) + \TVD(p_{\theta}, \phat_{\theta, n})\right]
        \nonumber.
    \end{IEEEeqnarray}
    The upper bound uses the decomposition and is a direct consequence of the triangle inequality. Specifically,
    \begin{IEEEeqnarray}{rCl}
     &&
        \frac{1}{2}\sum_{x \in \mathcal{X}}\sum_{y \in \mathcal{Y}}\Big|
            \Delta(x,y) + \Delta_{\theta, \ast}(x,y) + \Delta_{\theta, n}(x,y)
        \Big|
        \nonumber \\
     & \leq &
        \frac{1}{2}\sum_{x \in \mathcal{X}}\sum_{y \in \mathcal{Y}}|
            \Delta(x,y)| + |\Delta_{\theta, \ast}(x,y)| + |\Delta_{\theta, n}(x,y)|
        \nonumber \\
     & = &
        \TVD(p_{\ast}, p_{\theta, \ast}) + \left[\TVD(p_n, p_{\ast}) + \TVD(p_{\theta, \ast}, p_{\theta, n})\right].
     \nonumber
     \end{IEEEeqnarray}
     Noting the form of the upper and lower bounds together with the fact that 
     \begin{IEEEeqnarray}{rCl}
         \TVD(p, p_{\theta}) & = &
         \mathbb{E}_{p^x}\left[ 
         \TVD(p^{y|x}, f_{\theta})
         \right],
         \nonumber
     \end{IEEEeqnarray}
     it becomes clear that finding concentration inequalities for both $\TVD(\phat_n, p)$ and $\TVD(p_{\theta}, \phat_{\theta, n})$ towards zero suffices to prove the desired result.
     By assumption, it also holds that the joint distributions $p$ and $\phat_n$ have an alphabet of size at most $K_x \cdot K_y$. Numerous exponential concentration inequalities apply to this setting. While stronger results are available for the case where $n$ is small relative to $K_x + K_y$ \cite{concentrationTVDDiscrete}, we rely on Theorem 2.1 of \citet{concentrationTVDDiscrete_older} for simplicity and since the rates in $n$ remain the same. The latter shows that
     \begin{IEEEeqnarray}{rCl}
         \mathbb{P}\left( 
            \TVD( p, \phat_n ) > \varepsilon
         \right)
         & \geq &
         \delta(n, \varepsilon, K_x+K_y) 
         \nonumber \\
         \delta(n, \varepsilon, K) & = &
         (2^K-2) \cdot e^{  -n \varepsilon^2 /2 }
         \nonumber
     \end{IEEEeqnarray}
     Further, we also have that for any $\theta \in \Theta$,
     \begin{IEEEeqnarray}{rCl}
        \TVD(p_{\theta}, \phat_{\theta, n})
        & = &
        \frac{1}{2}
        \sum_{x \in \mathcal{X}}
        \sum_{y \in \mathcal{Y}}
        \left|
            p^x(x) - \phat[x]_{n}(x)
        \right|
        f_{\theta}(y|x)
        \nonumber \\
        & = &
        \frac{1}{2}
        \sum_{x \in \mathcal{X}}
        \left|
            p^x(x) - \phat[x]_n(x)
        \right|
        \underbrace{
        \sum_{y \in \mathcal{Y}}
        f_{\theta}(y|x)
        }_{\leq 1}
        \nonumber \\
        & \leq &
        \TVD(\phat[x]_n, p^x).
        \nonumber
     \end{IEEEeqnarray}
     Since $\phat[x]_n$ and $p^x$ have an alphabet of size at most $K_x$, the same type of concentration inequality applies here, too so that
     \begin{IEEEeqnarray}{rCl}
         \mathbb{P}\left( 
            \TVD( p^x, \phat[x]_n ) > \varepsilon
         \right)
         & \leq &
         \delta(n, \varepsilon, K_x) 
         \nonumber 
     \end{IEEEeqnarray}
     Setting now $\varepsilon = 2/\eta$ and using a union bound argument, we find that
     \begin{IEEEeqnarray}{rCl}
         && \mathbb{P}\left( 
            \TVD( p, \phat_n ) +
            \TVD( p^x, \phat[x]_n ) 
            > \varepsilon
         \right)
         \nonumber \\
         & \leq &
         \delta(n, 2/\eta, K_x) +
         \delta(n, 2/\eta, K_x + K_y)
         \nonumber \\
         & \leq &
         2\delta(n, 2/\eta, K_x + K_y)
         \nonumber 
     \end{IEEEeqnarray}
     Since by virtue of our previous derivations we also have that 
     \begin{IEEEeqnarray}{rCl}
     && \left|{\TVD}(\phat_n, \phat_{\theta, n})
     -
     \TVD(p, p_{\theta})\right|
     \nonumber \\
     & \leq &
     \TVD( p, p_n ) +
            \TVD( p^x, \phat[x]_n ),
     \nonumber
     \end{IEEEeqnarray}
     this completes the proof.
\end{proof}

\subsection{Proposition \ref{proposition:concentration_continuous}}

\begin{proof}
    The first part of the proof 
    follows exactly like in the discrete case. The only difference becomes the replacement of the summation $\sum_{x \in \mathcal{X}}$ with an integration operation. 
    Defining the joint, marginal and conditional kernel density estimates as
     \begin{IEEEeqnarray}{rCl}
         \phat[h_n]_n(x,y) & = & 
         \frac{1}{n}\sum_{i=1}^n \delta_{y_i}(y) \cdot \frac{1}{h_n^d}K\left(\frac{x_i - x}{h_n}\right) 
         \nonumber \\
         \phat[x,h_n]_n(x) & = &
         \sum_{y \in \mathcal{Y}}\phat[h_n]_n(x,y)
         \nonumber \\
         \phat[y,h_n]_n(y) & = &
         \int_{\mathcal{X}}\phat[h_n]_n(x,y)dx
         \nonumber \\
         \phat[x|y, h_n]_n(x|y) & = &
         \phat[h_n]_n(x,y) / \phat[y, h_n]_n(y)
         \nonumber \\
         p_n^{y|x, h_n}(y|x) & = &
         \phat[h_n]_n(x,y) / \phat[x, h_n]_n(x).
         \nonumber
     \end{IEEEeqnarray}
    Further, we define $\phat[ h_n]_{\theta,n}(x,y)  = f_{\theta}(x|y)\phat[x, h_n]_n(x)$ as the same hybrid-type distributions that were used in the proof for the discrete case.
    Using now the same basic inequalities as before, we find that 
    \begin{IEEEeqnarray}{rCl}
    &&
    \left|
    {\TVD}(p_{n}^{h_n}, \phat[ h_n]_{\theta,n})
    -
    \TVD(p, p_{\theta})
    \right|
    \nonumber \\
        & \leq & \TVD(p_n^{h_n}, p) + \TVD(p_{\theta}, \phat[ h_n]_{\theta,n}).
     \nonumber
     \end{IEEEeqnarray}
     Similarly, we can use the same arguments as in the discrete case to conclude that
     for any $\theta \in \Theta$,
     \begin{IEEEeqnarray}{rCl}
        \TVD(p_{\theta}, \phat[ h_n]_{\theta,n})
        & \leq &
        \TVD(\phat[x,h_n]_n, p^x).
        \nonumber
     \end{IEEEeqnarray}
     Notice that by definition, $\phat[x,h_n]_n$ is just a regular kernel density estimate for an absolutely continuous density.
     Thus, we can apply Theorem 1 (Chapter 3) in \citet{NonparametricDensityL1} to conclude that $\TVD(p_{\theta}, \phat[ h_n]_{\theta,n})$ goes to zero exponentially fast.

     Next, we use the triangle inequality to conclude that for $q(x,y) = \phat[x|y, h_n]_n(x|y)p^{y}(y)$,
     \begin{IEEEeqnarray}{rCl}
         \TVD(\phat[h_n]_n, p) 
         &  \leq  &
        \TVD(\phat[h_n]_n, q)
        +
        \TVD(q, p).
        \nonumber
     \end{IEEEeqnarray}
     Since it holds that
      \begin{IEEEeqnarray}{rCl}
         \TVD(\phat[h_n]_n, q) 
         &  =  &
         \frac{1}{2}
         \sum_{y \in \mathcal{Y}}
         \int_{\mathcal{X}}
         \phat[x|y, h_n]_n(x|y)
         \big|
         \phat[y, h_n]_n(y) - p^y(y)
         \big|dx
         \nonumber \\
         & = &
         \frac{1}{2}
         \sum_{y \in \mathcal{Y}}
         \big|\phat[y, h_n]_n(y) - p^y(y)\big|\underbrace{\int_{\mathcal{X}}\phat[x|y, h_n]_n(x|y)dx}_{= 1, \text{ for all } y}
         \nonumber \\
         & = &
         \TVD(
         \phat[y, h_n]_n, p^y).
         \nonumber
     \end{IEEEeqnarray}
     Notice that regardless of $h_n$, we actually have that $\phat[y, h_n]_n = \phat[y]_n = \frac{1}{n}\sum_{i=1}^n\delta_{y_i}(y)$, so that the same concentration inequalities in \citet{concentrationTVDDiscrete_older} already applied in the proof of the discrete case also hold for this last expression.
     For the second term resulting from applying the triangle inequality, we have that
     \begin{IEEEeqnarray}{rCl}
         \TVD(q, p)
         & = &
         \frac{1}{2}\sum_{y\in\mathcal{Y}}\underbrace{p^y(y)}_{\leq 1}\int_{\mathcal{X}}\big
         |\phat[x|y, h_n]_n(x|y) - p^{x|y}(x|y)\big|dx
         \nonumber \\
        & \leq &
         \sup_{y \in \mathcal{Y}} \TVD(\phat[x|y, h_n]_n(\cdot|y), p^{x|y}(\cdot|y))
         \nonumber.
     \end{IEEEeqnarray}
     Now note that by definition, for a fixed value of $y$, $\phat[x|y, h_n]_n$ is a regular kernel density estimate based on $n_y = \sum_{i=1}^n\delta_y(y_i)$ observations. 
     Thus, $\TVD(\phat[x|y, h_n]_n(\cdot|y), p^{x|y}(\cdot|y))$ satisfies a concentration inequality for any fixed $y$.
     We can use this knowledge as follows: Define the random variable $\widetilde{y}:\mathbb{N} \to \mathcal{Y}$ to be drawn uniformly from the set of maximizers
     \begin{IEEEeqnarray}{rCl}
         \arg\max_{y \in \mathcal{Y}}
         \TVD(\phat[x|y, h_n]_n(\cdot|y), p^{x|y}(\cdot|y))
         \nonumber
     \end{IEEEeqnarray}
     and note that by Theorem 1 in Chapter 3 of \citet{NonparametricDensityL1}\footnote{Though the constant is not stated explicitly in the original Theorem, one can work it out by tracing the relevant steps of Lemma 2 in the same Chapter and collecting the bounds.
     Doing so is tedious and yields a complicated expression which we have given an upper bound for here. } we have a \textit{conditional} upper-bound
     \begin{IEEEeqnarray}{rCl}
         \mathbb{P}\bigg(
            \max_{y \in \mathcal{Y}}
         \TVD(\phat[x|y, h_n]_n(\cdot|y), p^{x|y}(\cdot|y))
            > \eta
            \bigg| 
            \widetilde{y}(n) = z
            \bigg)
            &
            \leq 
            &
            \delta(z);
\nonumber
     \end{IEEEeqnarray}
     where for $n_z = \sum_{i=1}^n\delta_z(y_i)$ the number of times that $y_i = z$ occurred in the sample we have
     \begin{IEEEeqnarray}{rCl}
         \delta(z)
         & = &
         3e^{-n_z\cdot \eta^2 / 50} + e^{-2n_z\eta^2/25}.
         \nonumber
     \end{IEEEeqnarray}
     By taking $h:\mathcal{Y} \to [0,1]$ to be the true (and unknown) probability mass function of $\widetilde{y}$ and summing over the conditional upper bound, we obtain that 
     \begin{IEEEeqnarray}{rCl}
         && \mathbb{P}\bigg(
            \max_{y \in \mathcal{Y}}
         \TVD(\phat[x|y, h_n]_n(\cdot|y), p^{x|y}(\cdot|y))
            > \eta
            \bigg)
            \nonumber \\
            &
            \leq 
            &
            \sum_{z \in \mathcal{Y}}
            h(z)
            \delta(z)
            \leq \max_{z \in \mathcal{Y}} \delta(z);
            \nonumber \\
            & = & \delta_1 = 
            \min_{z \in \mathcal{Y}}\left\{
            3e^{-n_z\cdot \eta^2/ 50} + e^{-2n_z\eta^2 / 25}
         \right\}
         \nonumber
     \end{IEEEeqnarray}
     
     
     Based again on Theorem 1, Chapter 5 in \citet{NonparametricDensityL1}, one also finds that for a kernel density estimate $\phat[x, h_n]_n$ based on $n$ data points, 
     \begin{IEEEeqnarray}{rCl}
         \mathbb{P}\left(
            \TVD(\phat[x, h_n]_n, p^x)
            > \eta
         \right)
         \leq 
         \delta_2 
         =
         3e^{-n\cdot \eta^2 / 50} + e^{-2n\eta^2/25}.
         \nonumber
     \end{IEEEeqnarray}
     %
     Reusing the discrete concentration inequalities from before, we also have that
     \begin{IEEEeqnarray}{rCl}
        \mathbb{P}\left(
        \TVD(\phat[y, h_n]_n, p^y) > \eta
        \right)
        \leq 
        \delta_3
        =
        (2^{K_y}-2) \cdot e^{  -n \cdot \eta^2 /2 }.
        \nonumber
     \end{IEEEeqnarray}
     To obtain the union bound (and thereby the desired result), we now set $\eta = \varepsilon / 3$ to conclude that 
     \begin{IEEEeqnarray}{rCl}
         &&
         \mathbb{P}\left(
            \left|
    {\TVD}(\phat[h_n]_{n}, \phat[h_n]_{\theta, n} )
    -
    \TVD(p, p_{\theta})
    \right| > \varepsilon
         \right)
         \nonumber \\
         & \leq & \delta_1 + \delta_2 + \delta_3
         \nonumber \\
         & = &
         \min_{z \in \mathcal{Y}}\left\{
            3e^{-n_z\cdot \varepsilon^2/ 450} + e^{-2n_z\varepsilon^2 / 225}
         \right\}
         \nonumber \\
         && \quad
         +
         3e^{-n\cdot \varepsilon^2 / 450} + e^{-2n\varepsilon^2/225}
         +
         (2^{K_y}-2) \cdot e^{  -n \cdot \varepsilon^2 /18},
         \nonumber
     \end{IEEEeqnarray}
     %
     %
     which completes the proof.
\end{proof}

\subsection{Proposition \ref{proposition:consistency} }
\label{app:consistency}

\begin{assumption}
    $\theta^\ast$ is unique and or some $\varepsilon>0$,
    \begin{IEEEeqnarray}{rCl}
        B_{\varepsilon} & = &
        \left\{
            \theta: \left|
                \TVD(p, p_{\theta})
            - 
                \TVD(p, p_{\theta^\ast})
            \right| < \varepsilon
        \right\}
        \nonumber 
    \end{IEEEeqnarray} 
    is compact. Further, $\TVD(p, p_{\theta})$ and $\TVD(\phat_{n}, \phat_{\theta, n})$ are continuous on $B_{\varepsilon}$. Lastly,
    $\TVD(\phat_{n}, \phat_{\theta, n})$ has finite gradients (with respect to $\theta$) on $B_{\varepsilon}$ for all $n$ large enough in the sense that the finiteness condition
    \begin{IEEEeqnarray}{rCl}
    \limsup_{n\to\infty}
        \sup_{\theta' \in B_{\varepsilon}} \frac{\partial}{\partial\theta}\TVD(\phat_{n}, \phat_{\theta, n})\big|_{\theta = \theta'} < \infty
        \nonumber
    \end{IEEEeqnarray}
    holds almost surely.
    \label{assumption:consistency}
\end{assumption}

\begin{proof}
    We give the proof for the discrete case only, as the continuous case follows similar arguments.
    Roughly speaking, the proof proceeds in three steps: 
    First, we show that for large enough $n$, $\theta_n \in B_{\varepsilon}$. As this implies that we can confine the analysis to a compact set, it drastically simplifies the subsequent analysis.
    Second, we prove that over $B_{\varepsilon}$, $|\TVD(\phat_{n}, \phat_{\theta,n}) - \TVD(p, p_{\theta})|$ converges almost surely and \textit{uniformly} to zero. 
    Third, we use a standard argument to conclude that $\theta_n \overset{a.s.}{\to} \theta^{\ast}$.

    First, observe that
    \begin{IEEEeqnarray}{rCl}
        \TVD(p, p_{\theta_n}) 
        & \leq & 
        \TVD(\phat_{n}, p_{\theta_n}) + 
        \TVD(\phat_n, p)
        \nonumber \\
        & \leq &
        \TVD(\phat_{n}, p_{\theta^{\ast}}) + 
        \TVD(\phat_n, p)
        \nonumber \\
        & \leq &
        \TVD(p, p_{\theta^{\ast}}) + 
        2\cdot\TVD(\phat_n, p),
        \nonumber
    \end{IEEEeqnarray}
    where the first and last lines follow by the triangle inequality, while the second line follows by definition of $\theta_n$ and $\theta^{\ast}$.
    Further, note that by definition of $\theta^{\ast}$,
    \begin{IEEEeqnarray}{rCl}
        \TVD(p, p_{\theta^{\ast}})& \leq & \TVD(p, p_{\theta_n}).
        \nonumber 
    \end{IEEEeqnarray}
    Combining these two inequalities,
    \begin{IEEEeqnarray}{rCl}
    0 \leq 
        \TVD(p, p_{\theta^{\ast}}) - \TVD(p, p_{\theta_n})
        \leq 
         2\cdot\TVD(\phat_n, p).
        \nonumber 
    \end{IEEEeqnarray}
    Applying Theorem 2.1 of \cite{concentrationTVDDiscrete_older}, we know that $\TVD(\phat_n, p)$ converges to zero in probability exponentially fast. By the Borell-Cantelli argument already used for the proof of Corollary \ref{corollary:as_convergence}, this implies that $\TVD(\phat_n, p)\overset{a.s.}{\longrightarrow} 0$.
    Hence, for any $\xi > 0$ there is $N$ so that for $n\geq N$ and almost surely, 
    \begin{IEEEeqnarray}{rCl}
    0 \leq 
    \TVD(p, p_{\theta_n})
    -
        \TVD(p, p_{\theta^{\ast}})
        \leq 
         \xi.
        \nonumber 
    \end{IEEEeqnarray}
    Choosing $\xi = \varepsilon$, we can conclude that for $n\geq N$, $\theta_n \in B_{\varepsilon}$ (almost surely).

    In the second step, we use the fact that we can restrict our analysis to $n\geq N$ (i.e., to $B_{\varepsilon}$) in order to prove uniform convergence. Recall that by Corollary \ref{corollary:as_convergence}, we have pointwise convergence: for each $\theta \in B_{\varepsilon}$, $|\TVD(\phat_{n}, \phat_{\theta,n}) - \TVD(p, p_{\theta})| \overset{a.s.}{\to} 0$. 
    Further, $\TVD(\phat_{n}, \phat_{\theta,n})$ is strongly stochastically equicontinuous for $\theta \in B_{\varepsilon}$. This follows by a standard argument using the Mean Value Theorem in conjunction with the finiteness assumption on its gradients. Specifically, one can use the reasoning outlined on p. 340 in \cite{Davidson1994}, eqs. (21.55) -- (21.57) to conclude that the function is strongly stochastically equicontinuous (s.s.e.) by Theorem 21.10 of the same text.
    By Theorem 21.8 of the same text, this together with pointwise convergence implies that $\sup_{\theta \in B_{\varepsilon}}|\TVD(\phat_{n}, \phat_{\theta,n}) - \TVD(p, p_{\theta})| \overset{a.s.}{\to} 0$.

    The third and last step now consists in showing that as desired, $\theta_n \overset{a.s.}{\to} \theta^{\ast}$.
    This follows immediately by applying a standard result, see e.g. Lemma 2 in \cite{UniformLLN}.
    Notice that we can apply this result in spite of $\Theta$ being non-compact: As we only care about the limit, we can re-cast the minimization as occuring over a compact space. In particular, 
    \begin{IEEEeqnarray}{rCl}
        \min_{\theta \in \Theta}\TVD(\phat_{n}, \phat_{\theta,n}) 
        &=&
        \min_{\theta \in B_{\varepsilon}}\TVD(\phat_{n}, \phat_{\theta,n})
        \nonumber
    \end{IEEEeqnarray} 
    for all $n\geq N$.
    
    The proof for the continuous case proceeds along the same lines once one replaces $\phat_n$ with $\phat[h_n]_n$. The only complication is the exponential concentration inequality: Instead of Theorem 2.1 of \cite{concentrationTVDDiscrete_older}, one needs to use the results of \cite{NonparametricDensityL1} together with the arguments made for the proof of Proposition \ref{proposition:concentration_continuous}.
\end{proof}

\section{Experimental Details}

Whenever the optimization of the BFGS algorithm did not converge, we excluded the resulting samples of the Algorithm. 
This happened only for the $\varepsilon$-contaminated Poisson model (due to the numerically instable double-exponential parameterization of $\lambda$) and affected a negligibly small number of samples.

\subsection{Evaluation criteria}
\label{app:evaluation_criteria}
In our simulated experiments we consider three e aluation criteria: 1.\ the absolute difference between our inferred parameters and the truth, 2.\ absolute error on a test datasets and 3.\ the predictive likelihood.

Taking $\theta^{(j)}_{k}$ to be the $j$-th sample on the $k$-th artificially generated training dataset or split, our plots of criterion 1 then show quantiles of 
\begin{IEEEeqnarray}{rCl}
    d_{i,k} &=&\frac{1}{B} \sum_{j=1}^B |\theta_{k}^{(j)} - \theta|.
\end{IEEEeqnarray}

Next, we show quantiles of absolute errors 
\begin{IEEEeqnarray}{rCl}
    e_{i,k} &=& \frac{1}{B} \sum_{j=1}^B | y_{i,k}^{(j)} - y_{i,k}|,
\end{IEEEeqnarray}
where $y_{i,k}$ is the outcome of the $i$-th observation in the $k$-th artificially generated dataset and $y_{i,k}^{(j)}$ is the expected value of the distribution $p_{\theta^{(j)}_{k}}(\cdot|x_{i,k})$.

Lastly, we compute the predictive likelihoods for all experiments and show quantiles of
\begin{IEEEeqnarray}{rCl}
    l_{k,i} & = &\frac{1}{B}\sum_{j=1}^B p_{\theta^{(j)}_{k}}(y_{i,k}|x_{i,k}),
    \nonumber 
\end{IEEEeqnarray}
where $(y_{i,k}, x_{i,k})$ is the $i$-th observation in the $k$-th artificially generated training dataset or split.

\subsection{Full simulation results}\label{apx:experiments}
\begin{figure*}
\centering
\includegraphics[width=\textwidth]{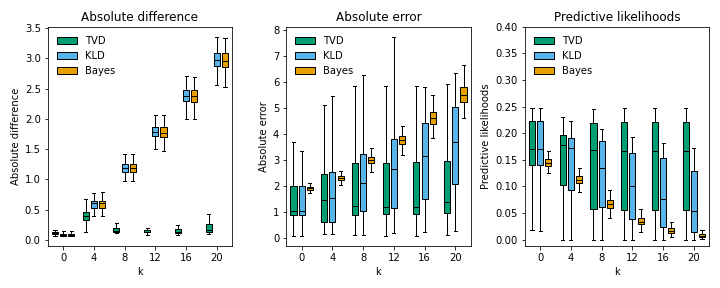}
\caption{Left: Absolute difference between estimated $\widehat{\lambda}$ and true data-generating $\lambda = 3$ for different values of $k$ with $\varepsilon = 0.15$ and $n_\text{train} = 400$. Middle: Absolute out-of-sample prediction error for different values of $k$ with $n_\text{test} = 100$. Right: Predictive likelihood on out-of-sample test data ($n_\text{test} = 100$) for different values of $k$.}
\label{fig:eps_stan}
\end{figure*}

\begin{figure*}
\centering
\includegraphics[width=\textwidth]{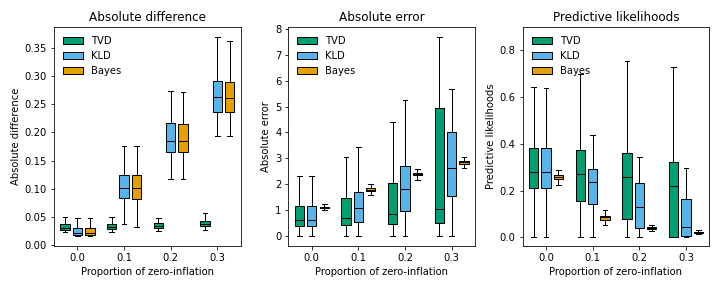}
\caption{Absolute difference between estimated $\widehat{\beta}$ and true data-generating $\beta = 0.25$ under increasing zero-inflation with $n_\text{train} = 800$. Absolute out-of-sample prediction error ($n_\text{test} = 200$) under increasing contamination. Predictive likelihood on out-of-sample test data ($n_\text{test} = 200$) with increasing contamination}
\label{fig:zero_stan}
\end{figure*}

In addition to performing inference with the \TVD and the \KLD reporting in \ref{sec:experiments} we also employed a fully Bayesian approach using Stan \cite{stan} with 4 chains and 1,000 posterior draws resulting in 4,000 draws from the posterior distributions in total. The full results are reported in Figures \ref{fig:eps_stan} and \ref{fig:zero_stan}. With respect to all three quantities of interest, the fully Bayesian approach is out-performed by the \TVD and the \KLD. This already demonstrates that using the Bootstrap approach proposed by \cite{Lyddon,EdwinNPL} entails a certain robustness because it incorporates uncertainty contained within the data. 

\subsection{Probit Models}

\begin{figure}[ht!]
\centering
\includegraphics[trim= {0.0cm 0.5cm 0.cm 10.68cm}, clip, width=\columnwidth]{probit_results.pdf}
\rule{\linewidth}{0.01cm}
\includegraphics[trim= {0.0cm 0cm 0.cm 10.68cm}, clip, width=\columnwidth]{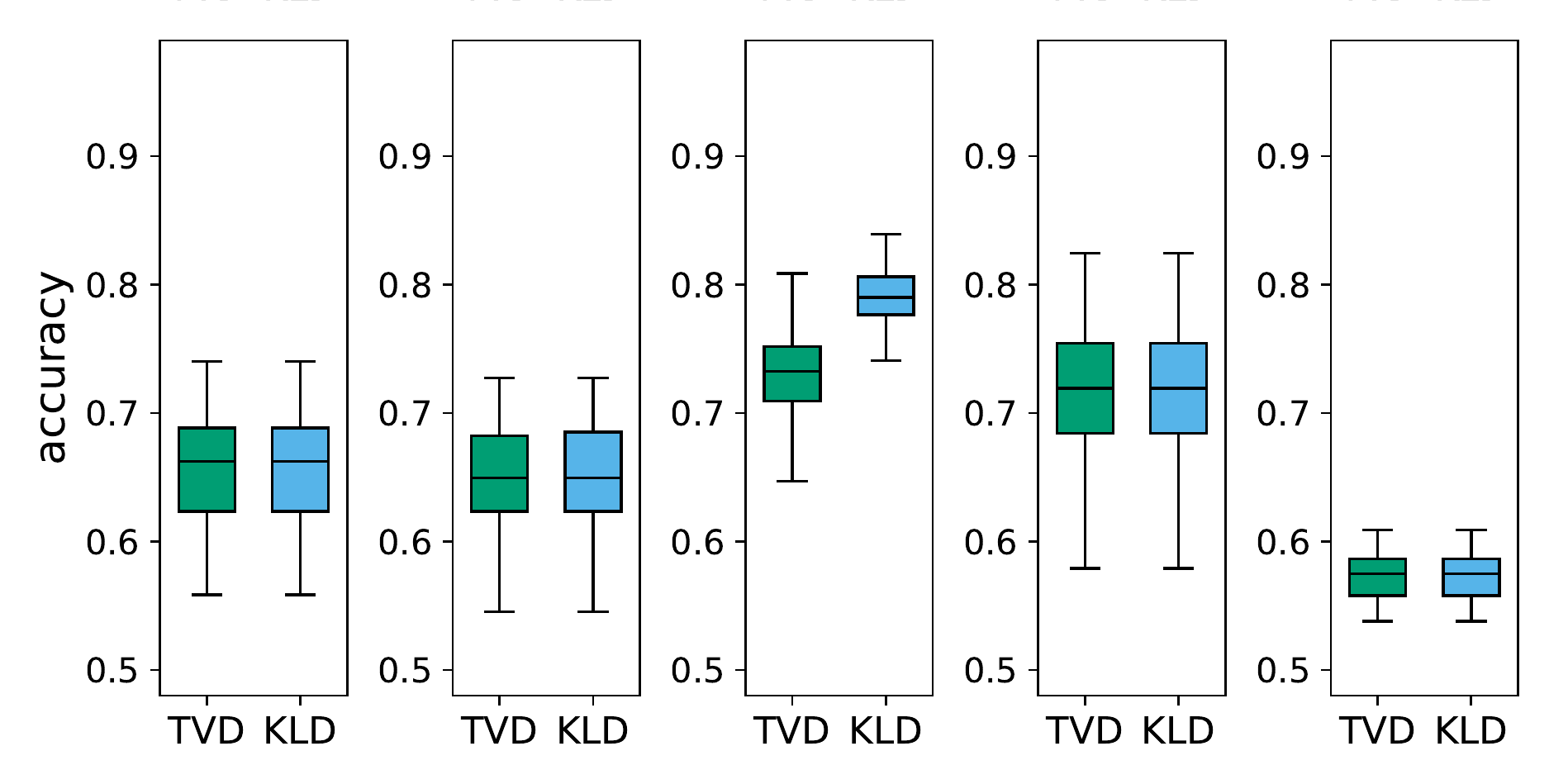}
\caption{Predictive accuracy from $50$ random splits for Probit models (top row) and Neural Networks (bottom row). }
\label{fig:probit_NN_results_appendix}
\end{figure}

For the \textit{fourclass} dataset, we test the first against the remaining classes.
For all datasets, we use {python}'s  \texttt{statsmodels} package to fit the (weighted) maximum likelihood estimate.
This estimate is the initializer for the BFGS algorithm used to minimize the \TVD.
Simultaneously, it is also a sample from the vanilla \NPL algorithm (where we minimize the negative log likelihood rather than the \TVD).
While using the \TVD yields a major improvement for the predictive likelihoods, the accuracy remains relatively similar to the \KLD case, see also the top row of Figure \ref{fig:probit_NN_results_appendix}. 

\subsection{Neural Network Models}

For simplicity, all Neural Network datasets are binary classification problems.
We use stochastic gradient descent within \texttt{Pytorch} \citep{pytorch} to get an initializer for the Neural Network. Once again, the BFGS algorithm is used to find the \TVD-minimizing value of $\theta$.
As for the Probit case, the accuracy remains relatively stable when the \KLD is replaced by the \TVD, see also the bottom row of Figure \ref{fig:probit_NN_results_appendix}.

\end{document}